\documentclass[gmd, manuscript]{copernicus}

\begin{document}
\nolinenumbers
\title{Representing Subgrid-Scale Cloud Effects in a Radiation Parameterization using Machine Learning: MLe-radiation v1.0}


\Author[1,2][hafner@iup.physik.uni-bremen.de]{Katharina}{Hafner} 
\Author[3]{Sara}{Shamekh}
\Author[4]{Guillaume}{Bertoli}
\Author[2]{Axel}{Lauer}
\Author[5]{Robert}{Pincus}
\Author[2]{Julien}{Savre}
\Author[2,1]{Veronika}{Eyring}

\affil[1]{University of Bremen, Institute of Environmental Physics (IUP), Bremen, Germany}
\affil[2]{Deutsches Zentrum für Luft- und Raumfahrt (DLR), Institut für Physik der Atmosphäre, Oberpfaffenhofen, Germany}
\affil[3]{Courant Institute of Mathematical Sciences, New York University (NYU), New York, NY, USA}
\affil[4]{Department of Earth and Environmental Engineering, Columbia University, New York, NY, USA}
\affil[5]{Lamont-Doherty Earth Observatory, Palisades, New York, USA}




\runningtitle{Improved radiation parameterization}

\runningauthor{K.~Hafner et al.}

\received{}
\pubdiscuss{} 
\revised{}
\accepted{}
\published{}


\firstpage{1}

\maketitle

\begin{abstract}
Improvements of Machine Learning (ML)-based radiation emulators remain constrained by the underlying assumptions to represent horizontal and vertical subgrid-scale cloud distributions, which continue to introduce substantial uncertainties. In this study, we introduce a method to represent the impact of subgrid-scale clouds by applying ML to learn processes from high-resolution model output with a horizontal grid spacing of $5\,\text{km}$. In global storm resolving models, clouds begin to be explicitly resolved. Coarse-graining these high-resolution simulations to the resolution of coarser Earth System Models yields radiative heating rates that implicitly include subgrid-scale cloud effects, without assumptions about their horizontal or vertical distributions. We define the cloud radiative impact as the difference between all-sky and clear-sky radiative fluxes, and train the ML component solely on this cloud-induced contribution to heating rates. The clear-sky tendencies remain being computed with a conventional physics-based radiation scheme. This hybrid design enhances generalization, since the machine-learned part addresses only subgrid-scale cloud effects, while the clear-sky component remains responsive to changes in greenhouse gas or aerosol concentrations. Applied to coarse-grained data offline, the ML-enhanced radiation scheme reduces errors by a factor of 4-10 compared with a conventional coarse-scale radiation scheme. This shows the potential of representing subgrid-scale cloud effects in radiation schemes with ML for the next generation of Earth System Models. 
\end{abstract}


\introduction  
For climate projections, coarse-scale Earth System Models (ESMs) typically have horizontal resolutions of $100-200\,\text{km}$ \citep{IPCC_2021_WGI_Ch_1}, in which clouds cannot be resolved explicitly. Therefore, these models require parameterizations of fractional cloudiness, particularly for cloud-radiation interactions at subgrid scales. A widely used approach is the Monte Carlo Independent Column Approach (McICA) \citep{Pincus2003}, where g-points are randomly assigned as cloudy or clear-sky according to the cloud fraction. This stochastic simplification introduces uncertainties in cloud-radiation interactions of up to  $100~W/m^2$ in surface fluxes, corresponding to relative errors of 10\% or more. However, these errors are unbiased compared to the Independent Column Approach (ICA) \citep{Barker1999}, where entire subcolumns are randomly designed as cloudy or clear, and thus tend to average out in long ESM simulations \citep{Pincus2003}. Vertical cloud overlap is commonly parameterized using the maximum-random overlap assumption \citep{Risnen2004}, whereby adjacent layers overlap maximally while distant cloud layers overlap randomly. Alternatively, some models employ an \textit{all-or-nothing} cloud cover scheme, which is a good approximation in high-resolution simulations \citep{Giorgetta2022,Hohenegger2023}.  

Machine learning (ML)-based radiation emulators have been developed for more than two decades \citep{ukkonen_exploring_2022,yao_physics-incorporated_2023,roh_evaluation_2020,pal_using_2019,krasnopolsky_new_2005,lagerquist_estimating_2023}. One potentially appealing aspect of ML-based emulators is their relative speed compared to traditional radiative transfer models, which in principle allows more frequent calls during ESM integration. In practice, however, the speed-up potential has proven limited \citep{Bertoli2025,ukkonen2025,hafner2025_stable}, with faster performance achievable through code optimization \citep{Cotronei2020,Ukkonen2024}. Moreover, improvements from more frequent radiation calls \citep{hafner2025_stable} remain marginal. Nonetheless, ML-based radiation emulators are valuable in specific context, such as differentiable ESMs \citep{kochkov_neural_2024,SpeedyWeatherJOSS}---where the derivative can be calculated for a complete model integration step---or GPU-based ESMs, where they are more energy efficient than physics-based schemes \citep{ukkonen2025}. Despite their high accuracy, emulators still face challenges in cloudy conditions \citep{hafner2024_interpretable}, which may partly explain why they have not yet shown substantial improvements over traditional radiation schemes. 

As more high-resolution Global Storm Resolving Model (GSRM) data become available, they offer increasing opportunities to enhance coarse-scale ESMs with machine learning. High-resolution simulations have been applied to nudge coarse-scale models toward fine-scale states \citep{Bretherton2022}, to learn subgrid tendencies directly \citep{Heuer2024,Busecke2025}, to infer subgrid effects from coarse-scale states \citep{grundner_deep_2022,Shamekh2023} and to learn all physics parameterizations \citep{WattMeyer2024}. Beyond these applications, high-resolution model simulations enable new strategies for representing fractional cloudiness and cloud overlap in coarse-scale ESMs using ML. Specifically, GSRM output can be coarse-grained to the resolution of the target ESM, and a neural network (NN) can then be trained to learn the subgrid-scale distribution of clouds from the underlying statistics. For instance, \citet{Henn2024} predicted coarse-grained cloud fields to reduce radiative biases; however, the cloud overlap assumption in the radiation scheme remains unchanged, limiting improvements. Leveraging ML to represent subgrid-scale cloud effects on radiation thus provide a promising pathway toward accurate radiation schemes in ESMs for climate projections. 

In this work, we demonstrate how the representation of cloud radiative impacts on heating rates can be improved by separating them from the all-sky heating rates and learning the cloud contribution directly from high-resolution simulations. Because these simulations explicitly resolve cloud systems without relying on assumptions about their horizontal and vertical distributions, the ML model can implicitly account for subgrid-scale cloud effects on the radiative heating rates. A similar separation between clear-sky and cloudy radiative fluxes was previously proposed by \citet{chevallier_neural_1998}. Moreover, \citet{meyer2022} focused on learning 3D cloud radiative effects. However, the explicit separation of cloud radiative impacts from all-sky radiation, especially including subgrid-scale effects, remains largely unexplored. 

We aim to use ML to encode the true variability of vertical and horizontal subgrid-scale clouds and their radiative impacts from high-resolution simulations, rather than relying on statistical schemes. This raises the central question: "Can the representation of cloud-radiation interactions be improved by training on high-resolution simulation data?" To address this, we compare our ML approach against a state-of-the-art physics-based radiative transfer model, namely RTE+RRTMGP \citep{pincus_balancing_2019} with McICA \citep{Pincus2003} and maximum-random overlap \citep{Risnen2004} to represent subgrid-scale cloudiness. In particular, we evaluate how a physics-based coarse-scale radiation scheme performs when applied to coarse-grained data. This comparison allows us to disentangle differences arising from changes in the input parameters from those due to different resolutions. 

This paper is structured as follows. Section \ref{sec:06:cre} describes the method used to learn the cloud radiative impact from high-resolution simulation data. Section \ref{sec:06:data} introduces the main datasets, generated with the ICOsahedral Non-hydrostatic (ICON) model\citep{giorgetta_icon-_2018, Giorgetta2022} in both high- and coarse-resolution configurations, and includes a comparison of the input and output variables of the radiation parameterization across these resolutions. In Section \ref{sec:06:pyrte_and ml_eval_on_coarse_grained}, we present the main results by comparing the physics-based coarse-scale radiation parameterization with the ML-enhanced scheme on coarse-grained test data. Finally, Section \ref{sec:06:conclusion} provides a discussion and conclusion.

\section{Learning the Cloud Radiative Impact on Heating Rates}
\label{sec:06:cre}

\begin{figure}[t]
    \centering
    \includegraphics[width=0.99\linewidth]{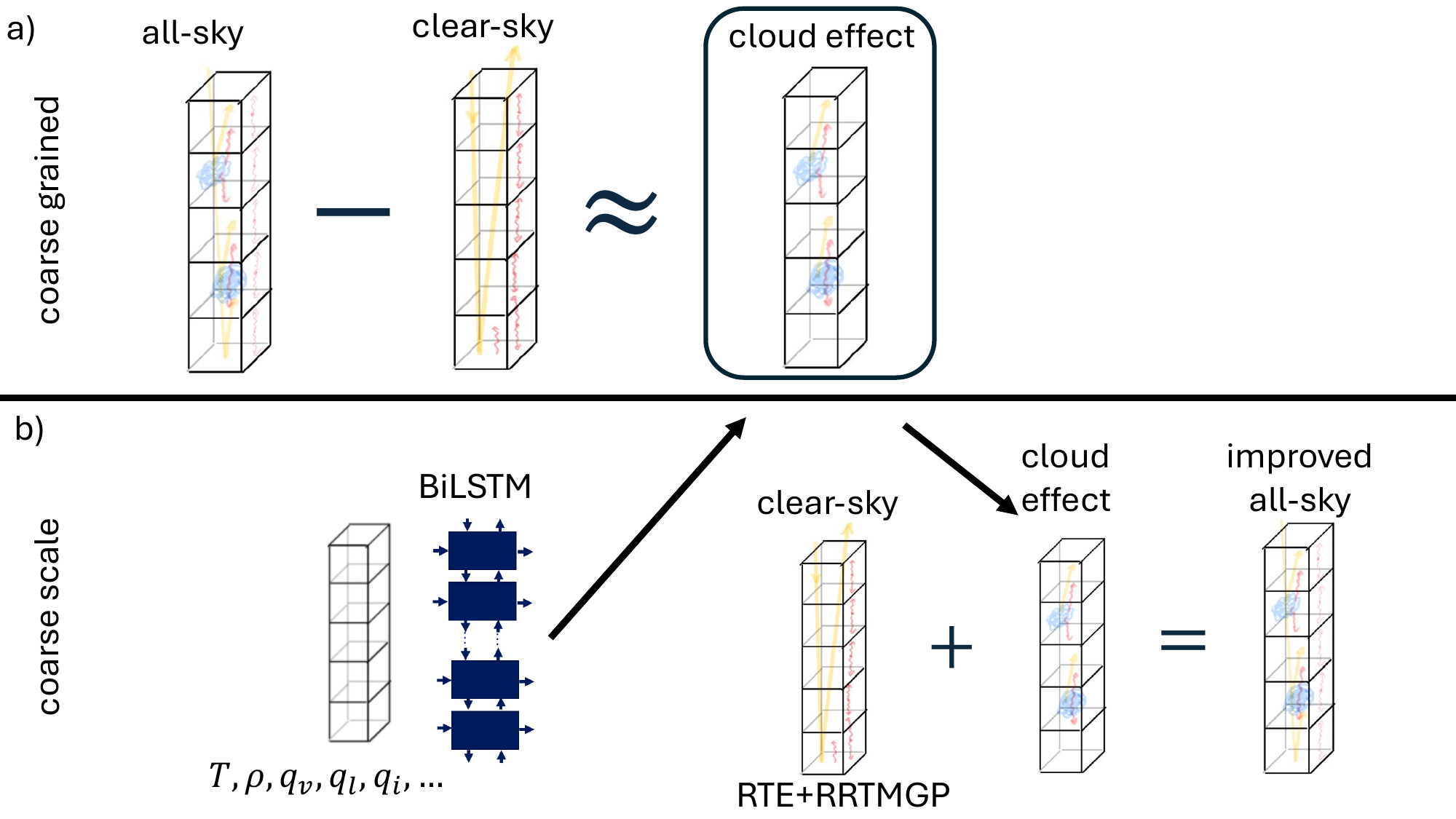}
    \caption[Sketch of learning the cloud radiative impact]{Sketch of constructing the cloud radiative impact on heating rates. Radiation schemes calculate fluxes for the same scene once with and once without clouds resulting in all-sky and clear-sky fluxes. The corresponding heating rates can be inferred from the fluxes and the residual yields an approximation of the cloud radiative impact on heating rates for all layers in a column.}
    \label{fig:06:cre_sketch}
\end{figure}
We define the cloud radiative impact on heating rates in a column as the residual between the all-sky and clear-sky heating rates, where clear-sky represents the same atmospheric conditions as all-sky but without clouds (Figure \ref{fig:06:cre_sketch} a). In ESMs such as the ICON model \citep{giorgetta_icon-_2018, Giorgetta2022}, radiation parameterizations like RTE+RRTMGP \citep{pincus_balancing_2019} first calculate gas optical properties and then add cloud optical properties. Then, these combined properties are used to calculate all-sky radiative fluxes, and clear-sky fluxes can be obtained omitting the cloud optical properties. Heating rates are then derived from the flux divergence for both all-sky and clear-sky conditions. The residual of these heating rates, which is the cloud radiative impact (CRI) on the heating rates, serves as the training target for the ML-based radiation parameterization:

\begin{equation}
    \frac{\partial T_{CRI}}{\partial t } = \frac{\partial T_{all-sky}}{\partial t } - \frac{\partial T_{clear-sky}}{\partial t}.
    \label{eq:CRI}
\end{equation}
The heating rates $\frac{\partial T_k}{\partial t}$ in a layer $k$ are calculated from the net flux $F_{Net}$ at the layer boundaries $k\pm\frac{1}{2}$ 
\begin{equation}
    \frac{\partial T_k}{\partial t} = \frac{F_{Net,k+1/2} - F_{Net,k-1/2}}{c_v m_{air}},
    \label{eq:06:conversion_mass}
\end{equation}
where $m_{air}$ is the mass of moist air per area, and specific heat at constant volume $c_v$ scales with the tracer mixing ratios.

The central idea is to train a neural network (NN) that learns only the cloud impact on heating rates (see Figure \ref{fig:06:cre_sketch}). Cloud-radiation interactions are subject to large uncertainties, since the subgrid-scale horizontal and vertical distributions of clouds are not resolved in coarse-scale ESMs. In the hybrid ML-physics radiation parameterization (Figure \ref{fig:06:cre_sketch} b), the NN predicts only the cloud radiative impact, while the clear-sky component is still computed by the original radiation parameterization. This design ensures that the ML-enhanced radiation scheme retains sensitivity to changes in GHG and aerosols through the clear-sky part, potentially improving generalization across different climates. The ML-cloud component can respond to GHG and aerosols changes only indirectly, for example through modifications of the cloud distribution. However, this hybrid approach does not capture secondary effects arising from reflected radiation. 

At first glance, a linear decomposition of the clear-sky heating rate and the cloud radiative impacts on heating rates may seem counterintuitive, given the inherently nonlinear interaction among specific humidity, clouds, and trace gases such as ozone \citep{Bony2015}. Nevertheless, we adopt a linear decomposition framework in this work, as illustrated on Figure \ref{fig:06:cre_sketch}. Within this framework, the NN is tasked with learning the nonlinear relationships based on the prevailing atmospheric conditions and cloud-related variables (e.g., cloud liquid water and cloud ice). 

\subsection{Method}
\label{sec:06:method}
We use a bidirectional long short term memory (BiLSTM) based on \citet{hafner2024_interpretable} to learn the cloud radiative impact in an atmospheric column. Bidirectional architectures are particularly well-suited for radiative transfer problems \citep{ukkonen_exploring_2022,yao_physics-incorporated_2023,ukkonen2025,Bertoli2025}. Unlike their common usage for temporal sequences, BiLSTMs for radiation scan the vertical dimension in both directions, resembling upward and downward fluxes. The NN consists of one BiLSTM layer with $tanh$ activation and a hidden dimension of 96, with two LSTMs scanning the input vertical profiles in upward and downward directions. The combined output of the LSTMs is then processed by a dense layer that predicts the heating rate at each level. The training is split between shortwave (SW) and longwave (LW) radiation as shortwave temperature tendencies are only calculated for sun-lit areas and longwave temperature tendencies are always computed, totaling in 82k trainable parameters per NN. 

The input variables are vertical profiles of mass mixing ratios of specific humidity $q_v$, cloud liquid $q_l$, cloud ice $q_i$, snow $q_s$ and ozone $O_3$, plus the cloud fraction $cl$, air density $\rho$, and temperature $T$. For SW, we additionally use surface albedo $\alpha$ and incoming solar flux at the top of the atmosphere $F_{\downarrow,TOA,SW}$, which is the solar constant weighted by the solar zenith angle and change in Earth-Sun distance to account for daily and seasonal variations. For LW, we additionally use the surface temperature $T_{surf}$ as input. As mentioned above, the output is the cloud radiative effect on heating rates derived from all-sky and clear-sky heating rates (Figure ~\ref{fig:06:cre_sketch}). 

Normalization is important for faster convergence of the training \citep{LeCun2012} and generalization \citep{beucler_climate-invariant_2024}. $O_3$, $\rho$, $T$, and $T_{surf}$ are normalized using their mean values $\mu$ and standard deviation $\sigma$, also known as z-score normalization ($x_{norm} = \frac{x-\mu}{\sigma}$). The water related variables $q_l$, $q_i$, $q_s$ are normalized by the ambient total (radiativly active) water ($q_v+q_l+q_i+q_s$). Here, radiatively active means used in the radiation scheme. $q_v$ uses z-score normalization providing information that relates to the absolute mass mixing ratios. $cl$ and $\alpha$ are not normalized as they already vary between 0 and 1. $F_{\downarrow,TOA,SW}$ is normalized by $1360~\frac{W}{m^2}$, which is close to the solar constant. The cloud radiative impact on heating rates is only converted to K/d, which is on the order of one. 

The loss we minimize during training consists of the sum of mean absolute error (MAE) and mean squared error (MSE). We use the Adam optimizer \citep{kingma_adam_2017}, and set a learning rate of $10^{-3}$, which is reduced by a factor of 2 when the validation loss does not decrease by 0.1\% for 5 epochs. To avoid overfitting, we use early stopping, which stops the training if the validation loss does not decrease for 10 epochs. 

\section{Data}
\label{sec:06:data}
ICON is a weather and climate model permitting simulations across different resolutions, from a few to hundreds of kilometers. For global and long-term applications, the atmospheric component ICON-A \citep{giorgetta_icon-_2018} often runs at $80\,\text{km}$ horizontal resolution, and 47 vertical levels covering the altitude range $0-83\,\text{km}$, with parameterizations for radiation, cloud microphysics, turbulence, convection and gravity waves. The ICON model has the option to be run as a GSRM \citep{Stevens2019,Giorgetta2022,Hohenegger2023}. The high-resolution simulations used here follow the QUBICC (Quasi-Biennial oscillation in a changing climate) protocol from \citet{Giorgetta2022}. The QUBICC simulations have $5\,\text{km}$ horizontal resolution, and the vertical dimension spans $83\,\text{km}$ on 191 levels. The high-resolution allows the model to run without a convection scheme and gravity wave parameterization, as these processes are starting to be resolved.

We performed QUBICC simulations that cover a total of 40 days evenly distributed across four months: November 2004, January, April, July 2005. The simulations have a physics time step of $40\,\text{s}$ and a radiation time step of $6\,\text{min}$. These simulations are run with prescribed sea surface temperatures, sea ice concentrations, greenhouse gas concentrations but no aerosols. The outputs are saved every $192\,\text{min}$. The uneven output interval was chosen to cover a large variability of different solar zenith angles at different locations. The first 6 days of each 10-day period is used for training, the next 2 days for validation and the last 2 days for testing. Using QUBICC data to train our ML model for ICON-A requires coarse-graining the high-resolution QUBICC dataset. All variables are horizontally and vertically coarse-grained from high-resolution simulations as in \citet{grundner_deep_2022}. We discarded a few coarse-grained cells, e.g., if the surface height of the coarse-grained cell deviated by more than $0.5\,\text{m}$ from the coarse-scale surface height. Then, we randomly sampled 35k and 5k grid points per time step for the training and validation set. For the test set, we randomly selected 75k cells per time step for LW and 35k cells for SW. This yields 2.3 million training samples, 260k validation samples, 1.9 million test samples for shortwave and 4.2 million test samples for longwave radiation.

In order to evaluate the high-resolution data, we also made coarse-scale ICON-A simulations for the same periods and with a similar configuration at $80\,\text{km}$ horizontal resolution to compare differences in various distributions. The ICON-A simulations are based on the version 2.6.4 described in \citet{giorgetta_icon-_2018}, while the QUBICC simulations are based on the version icon-2024.10 \citep{IconRelease2024.10}. To make the coarse-scale simulation more comparable, we used the same microphysics scheme \citep{doms2011} and no aerosols. See Appendix for a comparison with the default microphysics scheme. Both simulations use the radiation scheme RTE+RRTMGP \citep{pincus_balancing_2019}. The physics and radiation time step is $6\,\text{min}$ for the ICON-A simulation.

The remaining differences between the coarse-scale ICON-A and high-resolution QUBICC simulation include the horizontal and vertical resolution, the higher temporal resolution of physical process in QUBICC, which is required due to the high spatial resolution. Moreover, QUBICC intends to resolve gravity waves and (deep) convection. Additionally, the radiation scheme RTE+RRTMGP \citep{pincus_balancing_2019} in QUBICC uses snow mixing ratio as an input. Other differences could be due to differences between the code versions and tuning, as we did not retune ICON-A. 

\subsection{Comparison of input and output variables}
If an ML-based scheme is trained on high-resolution simulations like QUBICC and the goal is to couple it to a coarse-scale model like ICON-A, one needs to ensure that the distributions of input and output variables match. Otherwise, the ML-based scheme could be faced with out-of-distribution samples, which can lead to errors that quickly build-up while the model is integrated \citep{Rasp2020}. Therefore, we analyze systematic differences between the coarse-scale ICON-A simulations and the high-resolution QUBICC simulations. This analysis is conducted by comparing the spatial and temporal means and spread in the input and output used and produced by the radiation parameterization. Specifically, we focus on $q_v$, $q_l$, $q_i$, $cl$, total cloud cover, $q_s$, as well as longwave and shortwave heating rates. For the comparison, we use all samples in the test set. The samples of both simulations cover the same time period which is November 2004, January, April, and July 2005. Comparing two simulations with different grids enables us to uncover systematic differences, with a focus on identifying larger variations.

\begin{figure}[t]
    \centering
    \includegraphics[width=0.99\linewidth]{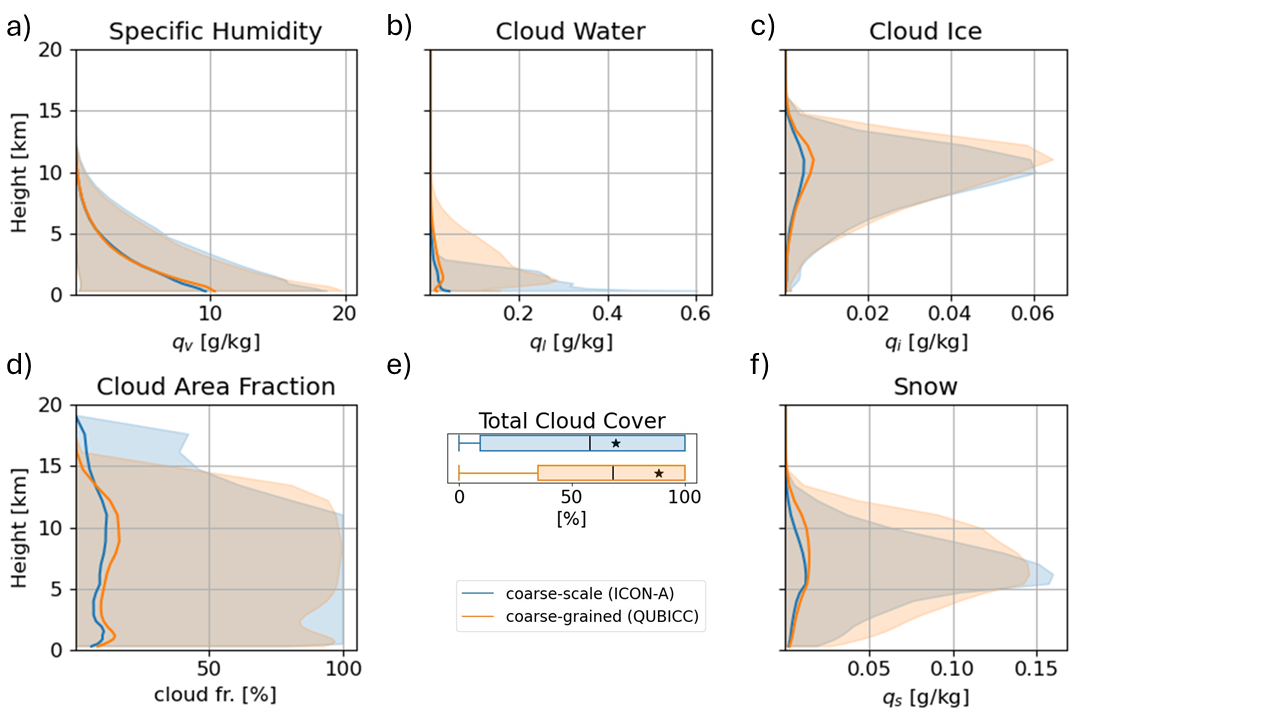}
    \caption[Input distribution coarse-scale vs coarse-grained]{Distributions of water related input variables for ICON-A and coarse-grained QUBICC data. The bold line shows the mean and the shaded area shows 95\% of the spread between the 2.5\% percentile and the 97.5\% percentile. The boxplot is limited by the minimum and maximum values. The box edges are defined at the 25\% and 75\% percentile of the distribution. The black line illustrates the mean of the distribution and the star is the median.   }
    \label{fig:06:input_distribution}
\end{figure}

Distributions of water related input variables are shown in Figure \ref{fig:06:input_distribution}. The distributions of specific humidity look similar in the coarse-grained and coarse-scale data set, where the maximum differences between the means is $0.9\,\text{g/kg}$ (Figure \ref{fig:06:input_distribution} a). The spread in humidity values also overlaps for both simulations. Cloud water has higher values below $3\,\text{km}$ for the coarse-scale simulation while cloud water is more evenly distributed throughout the troposphere for the QUBICC simulations (Figure \ref{fig:06:input_distribution} b). Here, the maximum difference of the mean values is $0.02\,\text{g/kg}$. The distributions of cloud ice have similar shapes, but the coarse-grained distribution peaks higher by $0.002\,\text{g/kg}$ (Figure \ref{fig:06:input_distribution} c). The mean cloud area fraction is on average larger by $3\,\%$ for the coarse-grained data set below $14\,\text{km}$ (Figure \ref{fig:06:input_distribution} d). Snow peaks between $5-10\,\text{km}$ (Figure \ref{fig:06:input_distribution} f). However, the vertical distribution is wider in the coarse-grained simulation. Nevertheless, snow was not used as input for radiation in ICON-A. 

\begin{figure}[t]
    \centering
    \includegraphics[width=0.99\linewidth]{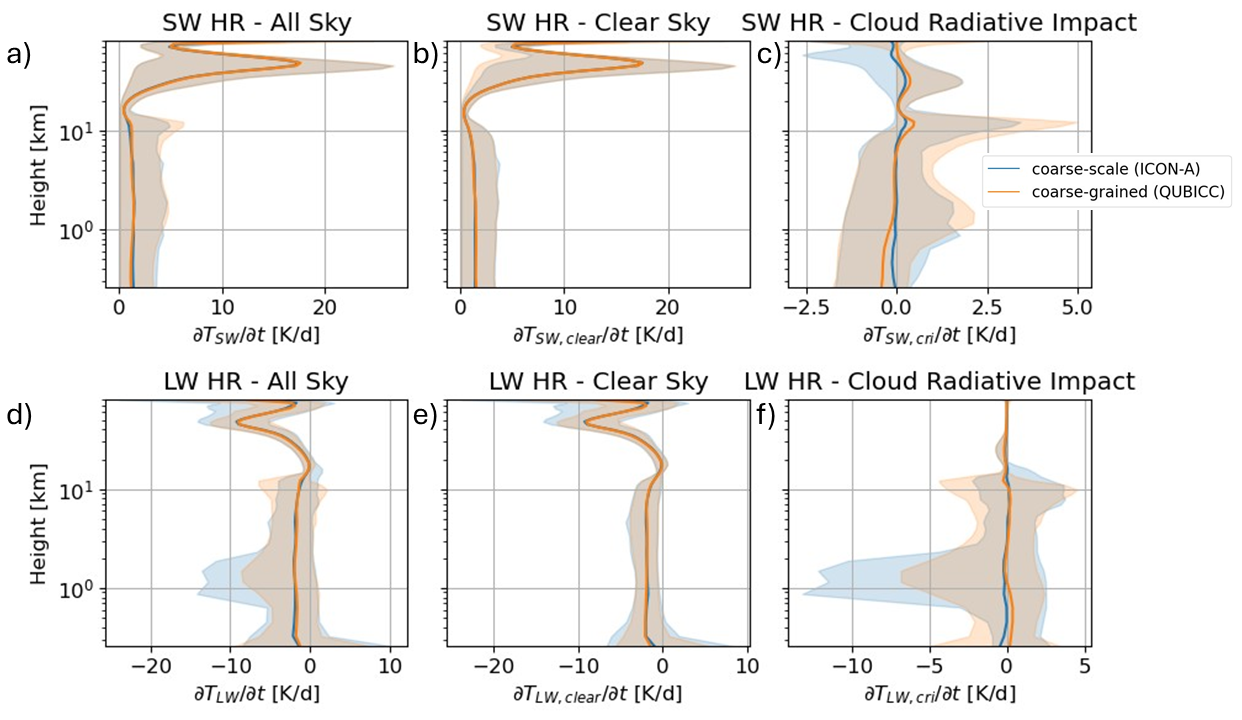}
    \caption[Output distribution coarse-scale vs coarse-grained]{The distribution of shortwave (top row) and longwave (bottom row) heating rates in coarse-scale and scaled coarse-grained data. The bold line shows the mean and the shaded area shows 95\% of the spread, which is defined as the spread between the 2.5\% percentile and the 97.5\% percentile. The left column shows all-sky heating rates as it is used in the ICON model. The middle column shows clear-sky heating rate computed from clear-sky fluxes, which is a diagnostic output in the ICON model. The right column shows the cloud radiative impact on the heating rate which was computed by subtracting the clear-sky heating rate from the all-sky heating rate.  }
    \label{fig:06:output_distributions}
\end{figure}

For comparability, the coarse-grained heating rates from QUBICC had to be rescaled by a factor $c_v/c_p$ to account for differences in the two model versions used, where $c_p$ is specific heat at constant pressure. 

For SW all-sky heating rates, the mean profiles match very well and the mean difference is $0.18\,\text{K/d}$ (Figure \ref{fig:06:output_distributions} a). There are only small differences in the spread of heating rates in the troposphere. The heating rate is decomposed into a clear-sky heating rate---which is calculated from the clear-sky fluxes---and the cloud radiative impact on heating rates. Their distributions are also shown in Figure \ref{fig:06:output_distributions} b-c. The SW clear-sky heating rate has a mean difference of $0.11\,\text{K/d}$ for the mean profiles. For the SW cloud radiative impact, the mean difference is also $0.11\,\text{K/d}$. Here, we expect small differences due to mostly resolved clouds in the coarse-grained dataset. However, there is no clear bias between coarse-scale and coarse-grained cloud impacts. 

For the LW all-sky heating rates, the mean profiles match well and have a mean difference of $0.18\,\text{K/d}$ (Figure \ref{fig:06:output_distributions} d). However, the spread in heating rates is slightly different, which may be due to the different spread in cloud water at $1\,\text{km}$ (Figure \ref{fig:06:input_distribution} b). The LW clear-sky heating rates look very similar in their mean values and their spread where the mean difference is $0.14\,\text{K/d}$ (Figure \ref{fig:06:output_distributions} e). The LW cloud impact is concentrated in the troposphere (Figure \ref{fig:06:output_distributions} f). The mean impact is almost the same between coarse-scale and coarse-grained simulations with a mean difference of $0.09\,\text{K/d}$. However, there is a difference in spread, which can be expected, because the coarse-grained data set implicitly includes subgrid-scale cloud effects.  

In the unscaled comparison, the clear-sky heating rates show the same bias as the all-sky heating rates for both SW and LW (Figure \ref{fig:C1:unscaled_output}). However, this bias does not directly translate to the cloud radiative impact on heating rates because adding the cloud impact is a highly non-linear process \citep{Bony2015}. Additionally, this indicates that the mean cloud effect is similar for a (quasi)-hydrostatic and a non-hydrostatic model.

\section{Results}
\label{sec:06:pyrte_and ml_eval_on_coarse_grained}
\begin{figure}[t]
    \centering
    \includegraphics[width=0.99\linewidth]{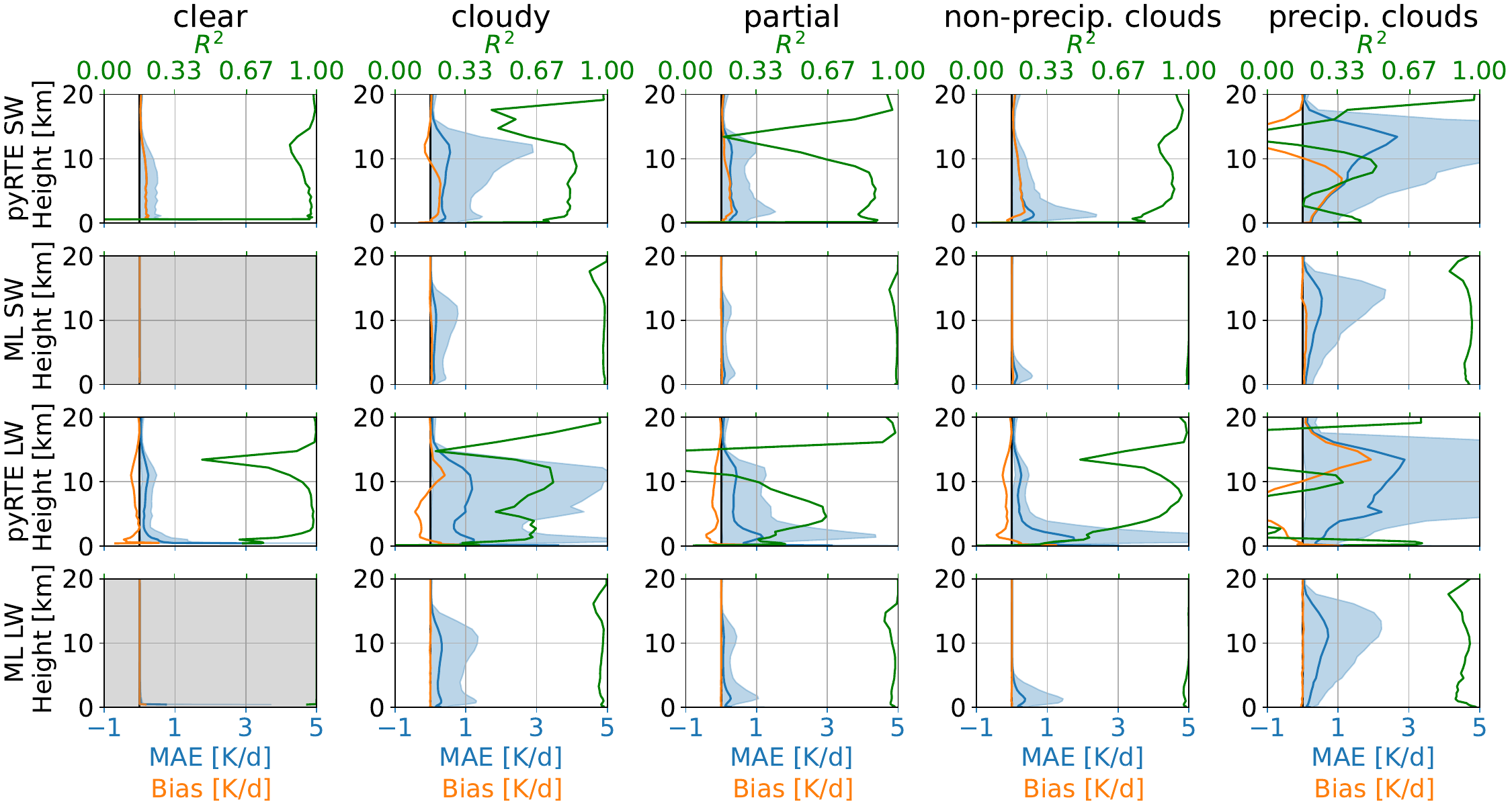}
    \caption[Comparison coarse-scale and ML-based radiation scheme on coarse-grained data]{Comparison of the pyRTE and the hybrid ML-based radiation scheme on coarse-grained QUBICC data. Results are shown for the shortwave (top rows) and longwave (bottom rows) spectral range. The results are shown separately for clear-sky samples (no clouds, left column), fully cloudy sky samples (middle column) and samples with partial cloudiness (right column). The shown metrics are coefficient of determination $R^2$ (green), bias (orange) and MAE (blue) with 95\% of the spread, which is defined as the spread between the 2.5\% percentile and the 97.5\% percentiles. The bias and MAE share the x-axis. The ML-clear-sky panels are gray because the clear-sky fluxes are not calculated by the ML model, but shown as reference. }
    \label{fig:06:comparison_by_clt}
\end{figure}

For the coarse-scale radiation reference, we use the Python front-end (hereafter pyRTE) \citep{Pincus_pyRTE-RRTMGP_2025} of the radiation scheme RTE+RRTMGP \citep{pincus_balancing_2019}, which is used in all our simulations. Using pyRTE, we replicate the implementation of radiation in QUBICC as closely as possible. For subgrid-scale variability, we employ McICA \citep{Pincus2003} together with maximum-random cloud overlap \citep{Risnen2004}. The procedure is as follows: first we calculate gas optical properties, assign them to the atmospheric state and calculate clear-sky fluxes. Next, we compute cloud optical properties, apply McICA with maximum-random overlap to represent subgrid-scale variability, and add cloud optical properties to the gas optical properties. Because QUBICC also includes snow in its radiation parameterization, we additionally calculate snow optical properties and combine them with the other optical properties, before computing all-sky fluxes. Then, heating rates are obtained applying Equation \ref{eq:06:conversion_mass}. We calculate heating rates for all samples in the test dataset and compare them to the coarse-grained heating rates. The same procedure is applied to the ML-enhanced heating rates, which are compared with pyRTE. As evaluation metric, we use mean absolute error (MAE), bias, and the coefficient of determination ($R^2$). The results are presented in Figure \ref{fig:06:comparison_by_clt} with the column-averaged metrics summarized in Table \ref{tab:C1:bulk_hr_clt}. We restrict the presentation to the lowest $20\,\text{km}$ of the atmosphere, as cloud impacts are most relevant in the troposphere (see Figure \ref{fig:06:output_distributions}). Nevertheless, the NN predicts the cloud impact on the entire atmospheric column, and the results for the full column can be found in the appendix \ref{fig:C1:full_column}.  

The first column of Figure \ref{fig:06:comparison_by_clt} shows clear-sky heating rates, for which the cloud distribution plays no role. Accordingly, only small and statistically insignificant differences are expected, arising mainly from variability in water vapor. For pyRTE, the assumption of horizontally homogeneous input parameters introduces a small error of $0.367\,\text{K/d}$ (SW) and $0.571\,\text{K/d}$ (LW). For comparison with other studies, Table \ref{tab:C1:bulk_hr_clt} also reports the root mean squared error (RMSE). For clear-sky heating rates from pyRTE, the RMSE is $0.443\,\text{K/d}$ (SW) and $0.688\,\text{K/d}$ (LW) compared with the coarse-grained QUBICC rates. \citet{hogan_tool_2022} developed a fast tool for computing gas-optical properties and reports an RMSE of $0.18\,\text{K/d}$. Although the error source differs (gas optical properties vs. spatial resolution), the magnitudes are comparable. 

For completeness, we also evaluated the ML-model on clear-sky samples. However, it is not intended for clear-sky scenes as the cloud impact is zero. Therefore, the corresponding panels are grayed out. The MAE is $0.049\,\text{K/d}$ for SW and $0.028\,\text{K/d}$ for LW. 

The second column of Figure \ref{fig:06:comparison_by_clt} shows results for fully cloudy samples (total cloud cover of 100\%). For pyRTE, the MAE peaks near $10\,\text{km}$, exceeding $5\,\text{K/d}$ for both SW and LW. The corresponding $R^2$ are low, with average values of 0.83 (SW) and 0.66 (LW), compared to 0.98 for the ML-enhanced scheme. $R^2$ is weighted by variability, and an $R^2$ of zero indicates that the error is as large as the variability itself. Since, McICA within pyRTE is supposed to produce unbiased noise, the $R^2$ is therefore a less informative metric. In contrast, the bias reveals that pyRTE systematically struggles to represent the cloud impact near $10\,\text{km}$, particularly for SW. The ML-enhanced scheme produces nearly unbiased heating rates, with MAEs of $0.106\,\text{K/d}$ (SW) and $0.127\,\text{K/d}$ (LW), representing errors 4-6 times smaller than those from pyRTE.

The third column of Figure \ref{fig:06:comparison_by_clt} shows results for partially cloudy scenes, defined here as total cloud cover between 10-90\%. In these cases, both schemes exhibit smaller errors than in fully cloudy scenes, consistent with the weaker overall cloud radiative impact. For pyRTE, the bias is substantially smaller than fully cloudy conditions but still show a pronounced peak near $1\,\text{km}$ for LW and a double peak at 1 and 10 km for SW. In contrast, the ML-enhanced samples produces nearly unbiased heating rates, with an MAE of $0.082\,\text{K/d}$ (SW) and $0.068\,\text{K/d}$ (LW), representing errors that are 5-10 times smaller than those for pyRTE. 

To interpret the double-peak error observed in partially cloudy samples, we divided the samples into precipitating and non-precipitating clouds, as a rough proxy for deep and shallow convection. Non-precipitating (warm) clouds were identified using thresholds of maximum $0.01\,\text{mm/h}$, total cloud cover larger than 10\%, and vertically integrated ice water path of less than $1\,\frac{g}{m^2}$. For reference, drizzle has a precipitation rate of $0.2\,\text{mm/d}$ \citep{Wood2012}. Precipitating clouds were identified by total cloud cover > 10\% and precipitation more than $3\,\text{mm/h}$. For reference, \citet{Zhao2024} reports mean precipitation of $3.5\,\text{mm/h}$ for deep convective cores over the tropical ocean. For non-precipitating clouds, both pyRTE and the ML-enhanced scheme show an MAE peak near $1\,\text{km}$ for SW and LW. However, the ML-enhanced scheme achieves substantially smaller errors of $0.080\,\text{K/d}$ for SW and $0.069\,\text{K/d}$ for LW, which are 6-10 times lower than those of pyRTE. For the precipitating clouds, pyRTE exhibits an MAE peak at $10-12\,\text{km}$, while the ML-enhanced scheme shows enhanced error in the upper troposphere but without distinct peak. Instead, the ML-enhanced scheme shows a broader peak in MAE between $12-14\,\text{km}$. On average, however, the ML-enhanced error remain about 4-5 times smaller than those from pyRTE. 

\begin{figure}[t]
    \centering
    \includegraphics[width=0.99\linewidth]{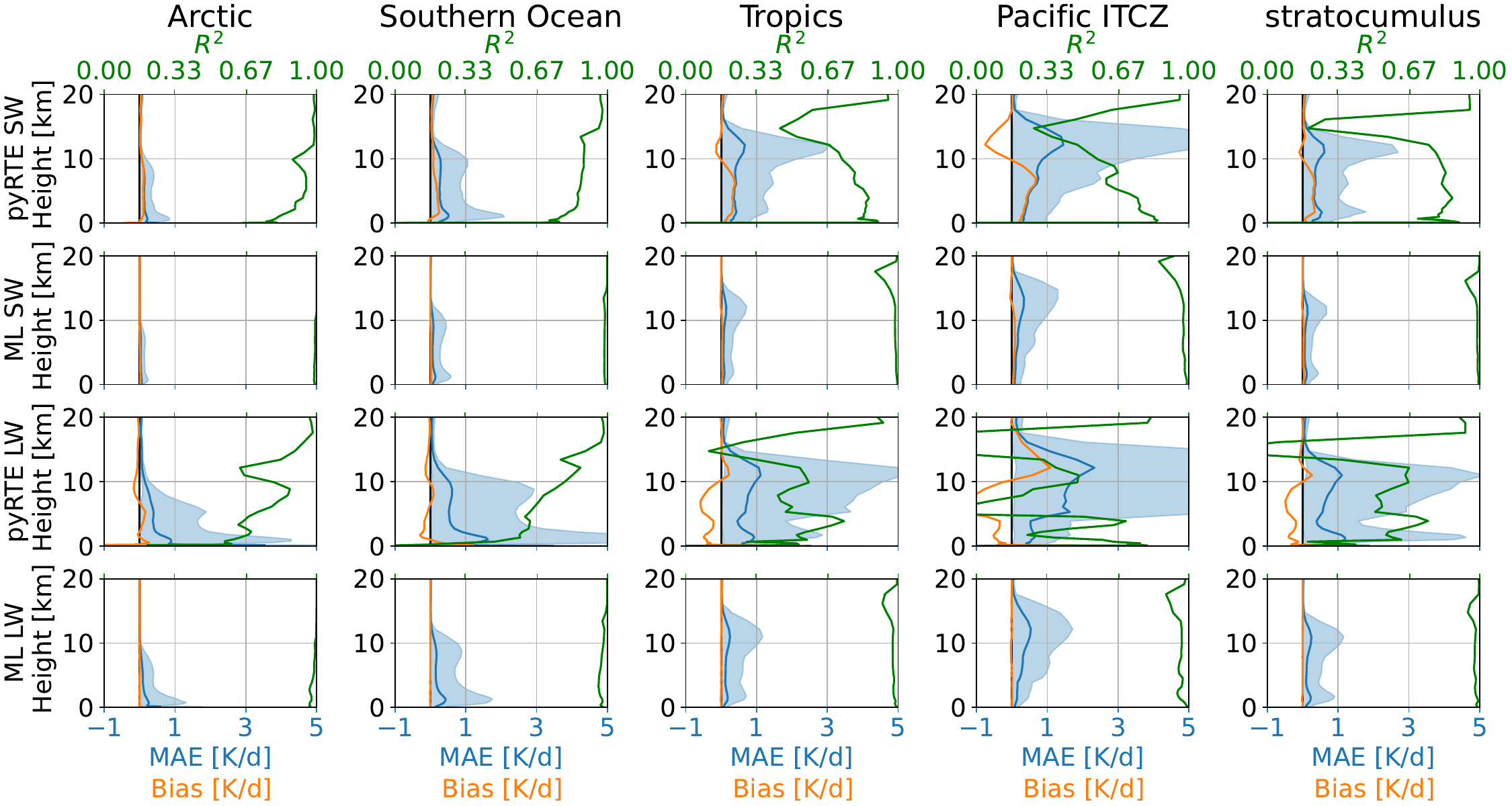}
    \caption[Comparison coarse-scale and ML-enhanced radiation scheme by region]{As Figure \ref{fig:06:comparison_by_clt}, but for the selected regions shown in Figure 5 of \citet{Bock2024}.}
    \label{fig:06:comparison_by_region}
\end{figure}

We further evaluated the performance of the ML-enhanced scheme across five selected regions characterized by different predominant cloud regimes, following the classification of \citet{Bock2024}. The results are shown in Figure \ref{fig:06:comparison_by_region} and summarized in Table \ref{tab:C1:bulk_hr_region}. In the arctic region (70-90\degree N), errors remain confined below $10\,\text{km}$ consistent with the lower tropopause height in this region. 
For pyRTE, the SW MAE is $0.215\,\text{K/d}$, even smaller than in the clear-sky conditions, although the spread in MAE is slightly larger. For LW, the MAE is $0.614\,\text{K/d}$, exceeding the clear-sky values. In contrast, the ML-enhanced scheme achieves errors that are 4-8 times smaller than those of pyRTE. 

In the Southern Ocean (30-65\degree S), pyRTE exhibits large errors of $0.417\,\text{K/d}$ for SW and $0.760\,\text{K/d}$ for LW, with a characteristic double peak in the MAE at 1-2 and 10\,km. The ML-enhanced scheme also reproduces this double peak structure in the MAE spread but reduces the MAE by a factor of 4-7 relative to pyRTE. Over the tropical ocean (30\degree N-30\degree S), pyRTE shows large errors in the upper troposphere, likely associated with deep convection. In this region, the ML-enhanced scheme again reduces the MAE by a factor of 5-9. A subregion within the tropical region, the Pacific ITCZ region (0-12\degree N, 135\degree E-85\degree W), shows similar behavior but with even larger errors at higher altitudes.

The stratocumulus region is represented by three sub-regions: the Southeast Pacific (10-30\degree N, 75-97\degree W), Southeast Atlantic (10-30\degree S, 10\degree W-10\degree E), and Northeast Pacific (15-35\degree N, 120-140\degree W). In these regions, both models show two peaks in the MAE: one in the lower troposphere at $1-2\,\text{km}$ and another in the upper troposphere at $10-12\,\text{km}$. Notably, the upper tropospheric peak is larger in both SW and LW for both models. Nevertheless, the ML-enhanced scheme achieves errors 5-9 times smaller than those from pyRTE.

\conclusions  
\label{sec:06:conclusion}

ESMs struggle to represent subgrid-scale cloudiness and commonly rely on statistical schemes, such as McICA, to account for subgrid-scale cloud radiative effects \citep{Pincus2003}. Although random, unbiased errors can be mitigated by large-scale atmospheric mixing \citep{Pincus2003}, the column-by-column error can be large. ML algorithms trained on high-resolution, global storm-resolving simulations now provide an opportunity to represent fractional cloudiness in radiative transfer more accurately. To bridge scales, the high-resolution model output is coarse-grained to the target resolution, such that the coarse-grained variables implicitly retain subgrid-scale effects. Then, the coarse-grained variables implicitly include subgrid-scale effects. 

We developed a hybrid physics-ML radiation parameterization, where the physics-based component computes clear-sky fluxes, while the ML component predicts cloud impact, implicitly accounting for subgrid-scale variability. This ML-enhanced framework offers a more robust and generalizable radiation scheme: the physics-based parameterization retains its responsiveness to changes in GHGs and aerosols, thereby mitigating potential out-of-distribution issues in climate projections. The ML component is implemented as a BiLSTM neural network, which has previously demonstrated strong performance in radiation applications \citep{ukkonen_exploring_2022,yao_physics-incorporated_2023,ukkonen2025,Bertoli2025,hafner2024_interpretable}. For training, we use data from high-resolution QUBICC simulations with a horizontal resolution of $\approx 5\,\text{km}$ and 191 vertical layers expanding up to $83\,\text{km}$. These fields are coarse-grained to $\approx 80\,\text{km}$ and 47 vertical layers, matching the target resolution for a coarse-scale ESM. For comparison and to assess systematic differences between high-resolution and coarse-scale models, we additionally perform a coarse-scale ICON-A simulation. The distributions of the relevant input variables are found to be comparable  between the coarse-scale and coarse-grained simulations.

We find that a coarse-scale radiation scheme such as pyRTE performs well for clear-sky samples, but exhibits large errors in cloudy conditions, reflecting its inability to represent subgrid-scale distributions. In contrast, the ML-enhanced radiation scheme consistently outperforms pyRTE, reducing errors from unresolved clouds in the radiative transfer calculations by a factor of 6-11. Although the ML-enhanced radiation scheme does not explicitly resolve subgrid-scale distributions, it learns how specific combinations of grid-scale mean states map to heating rates that implicitly include subgrid effects. pyRTE showed substantial biases of $1-3\,\text{K/d}$ in the upper troposphere at $10-15\,\text{km}$ for precipitating clouds, highlighting the strong influence of subgrid-scale cloud ice on heating rates. In general, both pyRTE and the ML-enhanced scheme produce larger errors in cloudier conditions, but the ML-enhanced scheme consistently yields smaller errors. These results emphasize the need for a more explicit treatment of subgrid-scale clouds, particularly in the upper troposphere. 

Therefore, we conclude that high-resolution model data combined with ML can improve the representation of cloud-radiation interactions in coarse-scale radiation parameterizations. Nevertheless, the presented approach has caveats. High-resolution simulations at $5\,\text{km}$ horizontal resolution cannot resolve shallow convection directly, leaving associated cloud radiative effects on heating rates---particularly within the planetary boundary layer---unresolved \citep{Stevens2019}. Using finer horizontal resolutions could help reduce these uncertainties. In addition, aerosols and heterogeneous GHG concentrations are typically absent from current high-resolution models. If future simulations include substantial variability in GHG and aerosol concentrations, these could be incorporated as additional NN inputs to capture secondary effects of reflected radiation.

One of the next steps is the online implementation of the ML-enhanced radiation scheme in a coarse-resolution model such as ICON-A. Although the online stability remains to be tested, the comparison with the coarse-scale model and results from previous stable hybrid simulations \citep{hafner2025_stable} are promising, suggesting potential improvement for climate projections. An additional advantage of the presented scheme is that clear-sky fluxes can be computed less frequently, while the cloud radiative impact can be updated every time step. This provides a pathway to both reducing computational costs and improving the representation of cloud-radiation interactions.


\codedataavailability{The software code for the ICON model is available at \url{https://icon-model.org} \citep{IconRelease2024.10} under the BSD-3C license. The exact version for high-resolution simulations is icon-2024.10 and for coarse-resolution, the exact version (2.6.4) corresponds to this commit: \url{https://gitlab.dkrz.de/icon/icon-model/-/tree/103515f9041eef54d6ff0290a758ee1da4d52f58}. The ICON-model repository is archived under \url{https://doi.org/10.35089/WDCC/IconRelease01} \citep{IconRelease2024.10}. The software code for pyRTE+RRTMGP is available \url{https://github.com/earth-system-radiation/pyRTE-RRTMGP} and archived under \url{https://doi.org/10.5281/zenodo.16644555} \citep{Pincus_pyRTE-RRTMGP_2025}. The code for the network training and plots is published at \url{https://github.com/EyringMLClimateGroup/hafner25GMD_MLe_radiation} and archived under \url{https://doi.org/10.5281/zenodo.17280639} \citep{code_mle_rad}.
} 

\appendix
\section{Default microphysics scheme} 
\label{sec:C1:microphysics}
\begin{figure}
    \centering
    \includegraphics[width=0.99\linewidth]{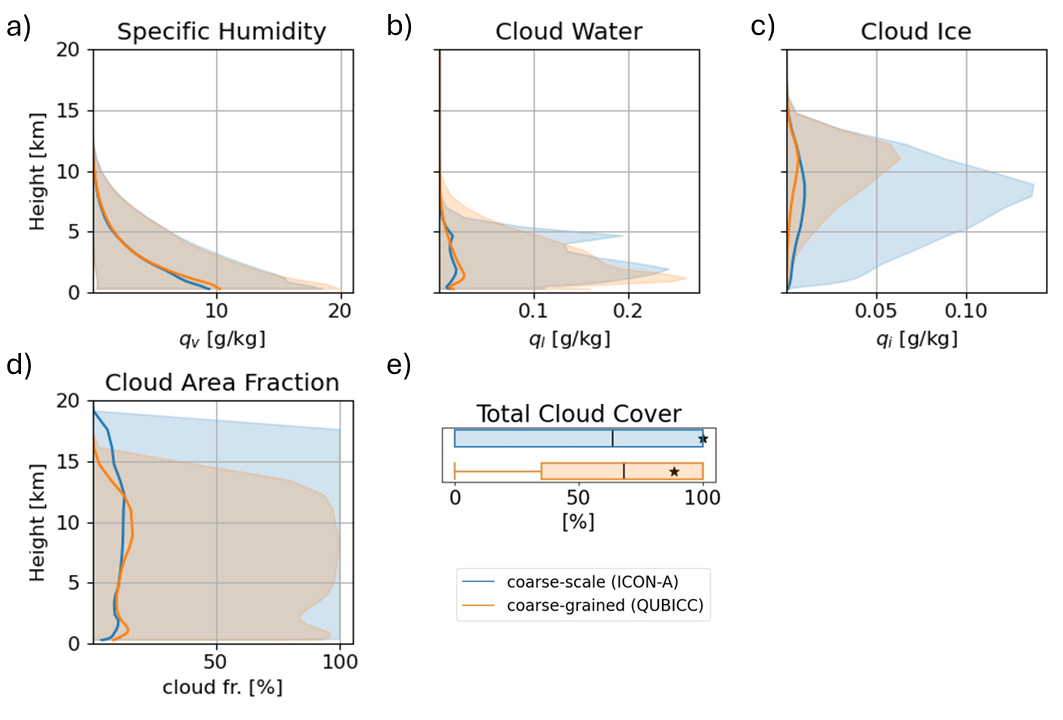}
    \caption[Input with default cloud microphysics scheme]{As Figure \ref{fig:06:input_distribution} but for the default microphysics scheme based on \citet{Lohmann1996}.}
    \label{fig:C1:input_lohman_roeckner}
\end{figure}

One major difference between the ICON-A and QUBICC versions is the microphysics scheme. ICON-A uses a modified version of the \citet{Lohmann1996} scheme and the QUBICC simulation uses the graupel scheme described in \citet{doms2011}. While both schemes are single-moment schemes, the latter treats precipitating tracers like snow, rain and graupel as prognostic variables while the former only diagnoses snow and rain. Moreover, it is known that cloud ice is too large in the upper troposphere in ICON-A \citep{Doktorowski2024}, which we also see when comparing cloud ice in ICON-A and QUBICC (Figure \ref{fig:C1:input_lohman_roeckner}).

\appendixfigures 

\section{Calculation of heating rates}    
\label{sec:C1:heaating_rates}
\begin{figure}
    \centering
    \includegraphics[width=0.99\linewidth]{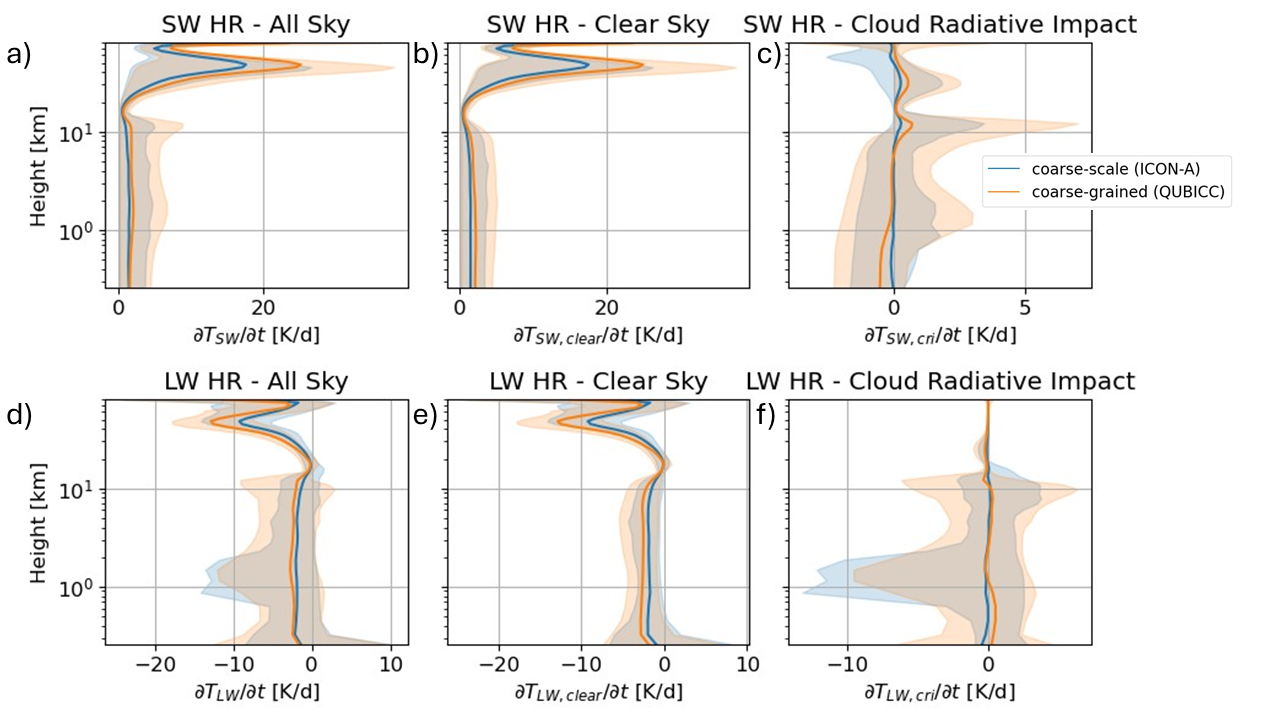}
    \caption[Unscaled heating rates coarse-scale vs coarse-grained]{As Figure \ref{fig:06:output_distributions} but here the coarse-grained heating rates are not scaled. }
    \label{fig:C1:unscaled_output}
\end{figure}

When comparing the unscaled heating rates between coarse-scale and coarse-grained simulations, we find huge biases of up to $10\,\text{K/d}$ between the mean heating rates, especially in the stratosphere. We found that there is a difference in heating rate calculation between the code versions. Usually, the heating rate is calculated from flux divergence, pressure difference and constants: 
\begin{equation}
    \frac{\partial T_k}{\partial t} = \frac{g}{c_p} \frac{F_{Net,k+1/2} - F_{Net,k-1/2}}{P_{k+1/2} - P_{k-1/2}},
    \label{eq:06:conversion_pressure}
\end{equation}
where $g$ is gravitational acceleration, $c_p$ specific heat at constant pressure, $F_{Net}$ is the difference between downward and upward flux, $P$ is pressure. $k$ is defined at the center of a layer (also full levels) while $k\pm\frac{1}{2}$ is defined at the layer boundaries (also half levels). This form of converting fluxes to heating rates is usually found in hydrostatic models but does not work in ICON because pressure is a diagnostic variable. Instead, the density is kept constant and specific heat at constant volume $c_v$ needs to be used for the conversion \citep{Zngl2014}. This transforms Equation \ref{eq:06:conversion_pressure} to Equation \ref{eq:06:conversion_mass}. In the ICON-A version, they use $c_p$ which is valid for quasi-hydrostatic models because then the hydrostatic pressure holds $dP/g = \rho dz$. Additionally, the specific heat is scaled only by water vapor, while all tracers are included in the code version used for QUBICC. Therefore, we scaled the coarse-grained heating rates in Figure \ref{fig:06:output_distributions} with the ratio $c_v/c_p$, which is on average 0.7. The unscaled heating rates are shown in Figure \ref{fig:C1:unscaled_output}.

\clearpage
\appendixtables   
\begin{table}[]
  \centering
  \caption[Results for heating rates of pyRTE vs ML-enhanced radiation scheme]{Bulk statistics for heating rate results of the coarse-scale ML-based radiation emulator on coarse-grained QUBICC data. MAE is mean absolute error and $R^2$ is coefficient of determination. RMSE is root mean squared error. The percentage values in parentheses denote the relative values of MAE, bias and RMSE. }
  \begin{tabular}{lcccc}
                    & MAE [$K/d$]    & Bias [$K/d$]     & $R^2$ & RMSE [$K/d$]\\
     \hline
     pyRTE &&&& \\
     \hline
     SW clear & 0.367 (8.47 \%) & 0.234 (4.91 \%) & 0.91 & 0.443 (10.56 \%) \\
     SW cloudy & 0.445 (16.73 \%) & 0.204 (4.22 \%) & 0.83 & 0.789 (41.54 \%) \\
     SW partial & 0.470 (12.24 \%) & 0.273 (4.65 \%) & 0.82 & 0.683 (23.55 \%) \\
     SW non-precip. clouds & 0.493 (12.11 \%) & 0.292 (4.79 \%) & 0.87 & 0.711 (19.98 \%) \\
     SW precip. clouds & 0.778 (32.62 \%) & 0.244 (17.28 \%) & 0.59 & 1.250 (58.35 \%) \\
     LW clear & 0.564 (23.56 \%) & -0.349 (-9.85 \%) & 0.83 & 0.677 (30.09 \%) \\
LW cloudy & 0.862 (41.08 \%) & -0.296 (-8.35 \%) & 0.67 & 1.478 (81.55 \%) \\
LW partial & 0.694 (25.83 \%) & -0.328 (-7.74 \%) & 0.56 & 1.112 (47.24 \%) \\
LW non-precip. clouds & 0.725 (48.35 \%) & -0.294 (-27.53 \%) & 0.70 & 1.230 (73.66 \%) \\
LW precip. clouds & 1.109 (77.76 \%) & -0.337 (-23.85 \%) & 0.34 & 1.578 (130.42 \%) \\
     \hline
     ML-enhanced &&&& \\
     \hline
     SW clear & 0.049 (0.49 \%) & -0.000 (-0.00 \%) & 0.99 & 0.074 (0.72 \%) \\
SW cloudy & 0.106 (4.46 \%) & 0.012 (0.98 \%) & 0.98 & 0.214 (11.30 \%) \\
SW partial & 0.082 (2.00 \%) & 0.005 (0.23 \%) & 0.99 & 0.150 (5.00 \%) \\
SW non-precip. clouds & 0.080 (1.57 \%) & 0.006 (0.24 \%) & 0.99 & 0.144 (3.50 \%) \\
SW precip. clouds & 0.188 (9.16 \%) & 0.024 (2.87 \%) & 0.96 & 0.341 (17.71 \%) \\
     LW clear & 0.028 (2.58 \%) & 0.003 (0.52 \%) & 1.00 & 0.053 (5.50 \%) \\
LW cloudy & 0.127 (7.42 \%) & -0.001 (0.00 \%) & 0.98 & 0.275 (17.70 \%) \\
LW partial & 0.068 (3.33 \%) & -0.001 (0.02 \%) & 0.99 & 0.158 (8.70 \%) \\
LW non-precip. clouds & 0.069 (4.43 \%) & -0.002 (0.43 \%) & 0.99 & 0.160 (9.41 \%) \\
LW precip. clouds & 0.197 (19.37 \%) & -0.004 (-0.99 \%) & 0.96 & 0.319 (36.18 \%) \\
  \end{tabular}
  
  \label{tab:C1:bulk_hr_clt}
\end{table}

\clearpage

\begin{table}[]
  \centering
  \caption[Results for heating rates of pyRTE vs ML-enhanced radiation scheme by region]{Bulk statistics for results of cloud radiative effect on heating rates. MAE is the mean absolute error and $R^2$ is the coefficient of determination. RMSE is the root mean squared error. The percentage values in parentheses denote the relative values of MAE, bias and RMSE. }
  \begin{tabular}{lcccc}
                    & MAE [$K/d$]    & Bias [$K/d$]     & $R^2$ & RMSE [$K/d$]\\
        \hline
        pyRTE &&&& \\
        \hline
        SW Arctic & 0.215 (9.55 \%) & 0.114 (3.73 \%) & 0.92 & 0.326 (17.99 \%) \\
        SW Southern Ocean & 0.417 (13.34 \%) & 0.218 (3.32 \%) & 0.87 & 0.665 (27.68 \%) \\
        SW Tropics & 0.535 (13.86 \%) & 0.290 (4.39 \%) & 0.80 & 0.817 (29.42 \%) \\
        SW Pacific ITCZ & 0.638 (18.33 \%) & 0.271 (6.27 \%) & 0.76 & 0.964 (33.67 \%) \\
        SW stratocumulus & 0.529 (13.66 \%) & 0.289 (4.38 \%) & 0.80 & 0.825 (28.97 \%) \\
        
        LW Arctic & 0.634 (30.75 \%) & -0.254 (-14.89 \%) & 0.77 & 1.046 (55.94 \%) \\
        LW Southern Ocean & 0.787 (30.60 \%) & -0.318 (-8.64 \%) & 0.69 & 1.307 (56.70 \%) \\
        LW Tropics & 0.755 (30.96 \%) & -0.326 (-4.21 \%) & 0.59 & 1.192 (65.09 \%) \\
        LW Pacific ITCZ & 0.944 (50.80 \%) & -0.311 (-2.19 \%) & 0.49 & 1.326 (99.80 \%) \\
        LW stratocumulus & 0.742 (31.38 \%) & -0.327 (-6.13 \%) & 0.58 & 1.234 (62.76 \%) \\
        \hline
        ML-enhanced &&&&\\
        \hline
        SW Arctic & 0.056 (2.17 \%) & 0.006 (0.58 \%) & 0.99 & 0.098 (4.84 \%) \\
SW Southern Ocean & 0.089 (2.98 \%) & 0.005 (0.34 \%) & 0.99 & 0.174 (7.30 \%) \\
SW Tropics & 0.097 (2.61 \%) & 0.009 (0.47 \%) & 0.98 & 0.187 (7.32 \%) \\
SW Pacific ITCZ & 0.142 (4.50 \%) & 0.019 (1.17 \%) & 0.98 & 0.247 (9.73 \%) \\
SW stratocumulus & 0.095 (2.48 \%) & 0.008 (0.41 \%) & 0.99 & 0.179 (6.38 \%) \\
        LW Arctic & 0.071 (4.96 \%) & -0.000 (0.01 \%) & 0.99 & 0.176 (11.91 \%) \\
LW Southern Ocean & 0.103 (4.86 \%) & -0.002 (-0.07 \%) & 0.99 & 0.231 (11.55 \%) \\
LW Tropics & 0.079 (4.67 \%) & -0.002 (-0.02 \%) & 0.99 & 0.172 (14.14 \%) \\
LW Pacific ITCZ & 0.132 (10.55 \%) & -0.002 (-0.46 \%) & 0.98 & 0.219 (23.47 \%) \\
LW stratocumulus & 0.080 (4.75 \%) & -0.001 (0.11 \%) & 0.99 & 0.175 (11.59 \%) \\
  \end{tabular}
  
  \label{tab:C1:bulk_hr_region}
\end{table}

\appendixfigures
\appendixfigures
\begin{figure}[t]
    \centering
    \includegraphics[width=0.99\linewidth]{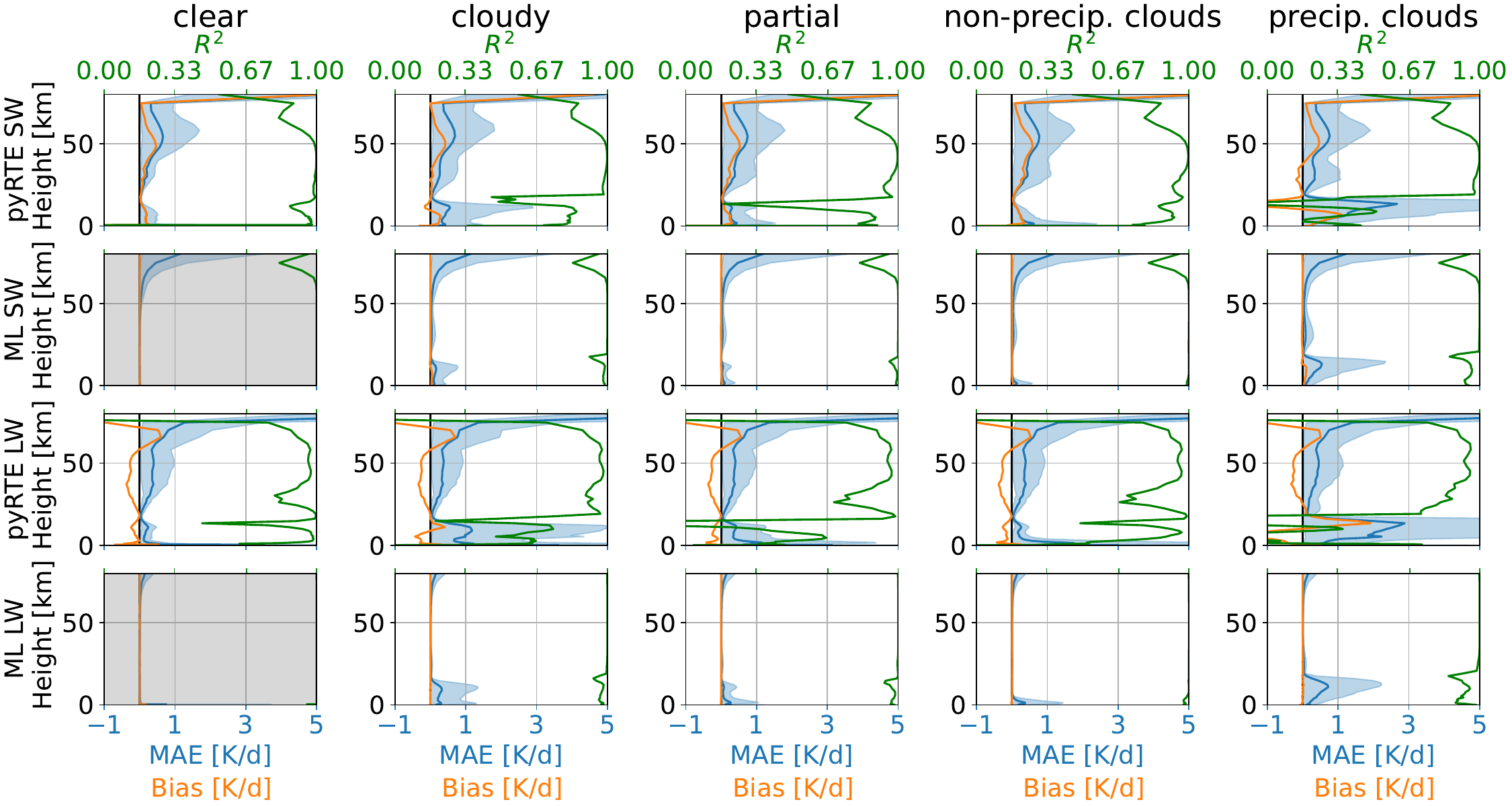}
    \caption[Comparison coarse-scale and ML-based radiation scheme on coarse-grained data]{As Figure \ref{fig:06:comparison_by_clt} but for the full column.}
    \label{fig:C1:full_column}
\end{figure}

\authorcontribution{
Katharina Hafner: Conceptualization, Investigation, Formal analysis, Software,  Data Curation, Methodology, Writing - Original Draft
Sara Shamekh: Conceptualization, Methodology, Writing - Review \& Editing
Guillaume Bertoli: Conceptualization, Methodology
Axel Lauer: Supervision, Writing - Review \& Editing
Robert Pincus: Conceptualization, Software
Julien Savre: Data Curation, Writing - Review \& Editing
Veronika Eyring: Funding acquisition, Supervision
} 

\competinginterests{Some authors are members of the editorial board of journal Geoscientific Model Development.} 


\begin{acknowledgements}
KH and VE were supported by the Deutsche Forschungsgemeinschaft (DFG, German Research Foundation) through the Gottfried Wilhelm Leibniz Prize awarded to Veronika Eyring (Reference No. EY 22/2-1). VE additionally acknowledge funding by the European Research Council (ERC) Synergy Grant ``Understanding and Modeling the Earth System with Machine Learning'' (USMILE) under the Horizon 2020 Research and Innovation program (Grant Agreement No. 855187). This work used resources of the Deutsches Klimarechenzentrum (DKRZ) granted by its Scientific Steering Committee (WLA) under project ID bd1179. RP, and SS were supported by the US National Science Foundation through the Learning the Earth with Artificial intelligence and Physics (LEAP) Science and Technology Center (STC) (Award \#2019625). SS acknowledges support provided by Schmidt Sciences, LLC. The authors gratefully acknowledge the Earth System Modelling Project (ESM) for funding this work by providing computing time on the ESM partition of the supercomputer JUWELS at the Jülich Supercomputing Centre (JSC).
\end{acknowledgements}



\bibliographystyle{copernicus}
\bibliography{paper.bib}

\end{document}